\documentstyle[psfig,prl,aps,twocolumn]{revtex}

\begin{document}
\draft

\title{A new family of models with exact ground states\\ 
connecting smoothly the $S=\frac{1}{2}$ dimer and $S=1$ Haldane phases 
of 1D spin chains}

\author{A. K. Kolezhuk\cite{perm1}}
\address{Institute of Magnetism, National Academy of
Sciences and Ministry of Education of Ukraine, 252142 Kiev, Ukraine\\
and Institut f\"{u}r Theoretische Physik,
Universit\"{a}t Hannover, 30167 Hannover, Germany}

\author{H.-J. Mikeska\cite{perm2}}
\address{Institut f\"{u}r Theoretische Physik,
Universit\"{a}t Hannover, 30167 Hannover, Germany\\
and Department of Earth and Space Science, University of Osaka,
Toyonaka, Osaka 560, Japan}
\date{\today}

\maketitle

\begin{abstract}
We investigate the isotropic two-leg $ S=\frac{1}{2}$ ladder with
general bilinear and biquadratic exchange interactions between spins on
neighboring rungs, and determine the Hamiltonians which have a matrix
product wavefunction as exact ground state. We demonstrate that a smooth
change of parameters leads one from
the $S=\frac{1}{2}$ dimer and Majumdar-Ghosh chains to the $S=1$ chain with
biquadratic exchange. This proves that these
model systems are in the same phase.
\end{abstract}


Low dimensional quantum antiferromagnets have attracted a large amount of both
theoretical and experimental interest in recent years. Theoretically, the
results of calculations for model systems in one dimension (1D) clearly show
the difference between systems with a gapless spectrum of excitations and
power law decay of spin correlations and gapped systems with exponentially
decaying correlation functions. The main representatives of these two classes
are resp. the $S=\frac{1}{2}$ chain with isotropic nearest neighbor (NN)
exchange interaction \cite{Bet31} and the isotropic $S=1$ (Haldane) chain
\cite{Hal83}. Gapped elementary excitation spectra are also found in more
complicated low-dimensional $S=\frac{1}{2}$ systems such as chains with
sufficiently strong additional next-nearest neighbor (NNN) exchange or with
alternating exchange (with noninteracting dimers as the simplest limit) and
also isotropic spin ladders with an even number of legs \cite{DagR96}.

Many of these models are realized to a high degree of accuracy in simple
compounds as demonstrated by the following examples: $\rm KCuF_3$ (1D
isotropic $S=\frac{1}{2}$ antiferromagnet
\cite{NagTCPS91}), $\rm CaCuGe_2O_6$  (weakly interacting dimers
\cite{ZheSSHU96}), $\rm (VO_2)P_2O_7$ \cite{EccBBJ94} and $\rm SrCu_2O_3$
\cite{AzuHTIK94}
(two leg spin ladder) and $\rm Ni(C_2H_8N_2)_2NO_2ClO_4$ (NENP, 1D $S=1$
antiferromagnet \cite{RenGRV90}).
Although the precise spin Hamiltonians of these substances are not known,
the study of the simple theoretical models mentioned above has contributed
in an essential way to the understanding of the behaviour of the real
materials and these are believed to be in the same phase as
the simple models.

In recent years arguments and numerical evidence have been presented that
the various models with a gapped excitation spectrum, in particular the
$S=\frac{1}{2}$ dimer chain, the gapped nearest and next nearest  neighbor
exchange chain, the two-legged $S=\frac{1}{2}$ spin
ladder and the $S=1$ (Haldane) chain are all in the same phase
\cite{KohT92,Hid91,KatI95,TotS95a,BreMN96,Whi96}. In this letter we present the
following proof for this conjecture: We describe a class of Hamiltonians
with exactly known nondegenerate ground states and show that it is
possible, by a continuous change of parameters within this class, to
connect smoothly the hamiltonians of the 1D dimer chain, the Majumdar-Ghosh
chain, a generalized spin ladder (which includes biquadratic interactions)
and of the $S=1$ antiferromagnetic 
chain with additional biquadratic exchange. Given the
relevance of these model systems with exactly known ground states for the
traditional model systems and for the real quasi 1D compounds this proves
the existence of a single phase for the gapped low dimensional
spin systems above. As a ``byproduct'' of this investigation, we also present
a new set of models of frustrated $S=\frac{1}{2}$ spin chains (including only
bilinear NN and NNN interactions) whose ground states can be found exactly. 

Our approach starts from the observation \cite{BreMN96} that the exactly known ground
states of
the Shastry-Sutherland $S=\frac{1}{2}$ chain \cite{ShaS81}
\begin{equation}
H= \sum_n  (1 + (-1)^n \delta) 
  {\mathbf S}_n {\mathbf S}_{n+1}
 +{1\over2} (1-\delta) {\mathbf S}_n {\mathbf S}_{n+2} ,
\label{shas}
\end{equation}
which includes the Majumdar-Ghosh ($\delta = 0$) \cite{MajG69}
and dimer ($\delta = 1$) limits as special cases, and the $S=1$ chain with
biquadratic exchange
\begin{equation}
H= \sum_n  {\mathbf S}_n {\mathbf S}_{n+1} 
- \beta ({\mathbf S}_n {\mathbf S}_{n+1})^2
\label{aklt}
\end{equation}
at $\beta = - \frac{1}{3}$ (the AKLT model \cite{AffKLT87} which is believed
to be in the same phase as the Haldane chain with $\beta=0$)
have very similar structure: both ground states can be written as matrix
product (MP) wave functions,
\begin{equation}
|\psi_0 \rangle = \prod_i  g_i,\;\;
g_i  =  \left( 
\begin{array}{lr}
b | s \rangle_i +a | t_0\rangle_i & - \sqrt{2} a | t_{+1} \rangle_i \\
\sqrt{2} a | t_{-1} \rangle_i & b | s \rangle_i - a | t_0 \rangle_i
\end{array} 
\right).
\label{matrix}
\end{equation}
The ground state of the Shastry-Sutherland chain is obtained setting
$a=b=\frac{1}{2}$, where $| s \rangle_i$ and $| \vec t \rangle_i$ are the
singlet, resp. triplet formed by two spins $S=\frac{1}{2}$ on adjacent sites
$n$ and $n+1$ ($n$ even for $\delta > 0$ in (\ref{shas})).  The ground state
of the AKLT chain is obtained by setting $a=\frac{1}{\sqrt{3}}$, $b=0$, and
identifying the remaining triplet with the three states of the spin $S=1$ at
site $i$. The matrix elements of the MP wavefunction $| \psi_0 \rangle$ are
related to boundary effects for open boundary conditions whereas for periodic
boundary conditions the trace should be taken \cite{AffKLT87}. Matrix product
(or finitely correlated) wave functions were first introduced by Fannes et al
\cite{FanNW89} a decade ago and have since then found widespread applications
in exact and variational calculations
\cite{KenT92,KluSZ91,NiggZitt96,GSu96,TotS95}. These wave functions are
characterized by finite (and typically rather short) correlation lengths and
are therefore tailored to deal with gapped systems.

It should be emphasized that our MP ansatz for the ladder (\ref{matrix}), in
contrast to that used in Ref.\ \onlinecite{GSu96}, respects rotational
symmetry of the Hamiltonian (\ref{ham}), i.e. the wavefunction
$|\psi_{0}\rangle$ under periodic boundary conditions is a global singlet for
any value of the parameter $u$
\cite{BreMN96,KolMY97}. The ansatz used by Su\cite{GSu96} has a `built-in'
broken rotational symmetry and thus can access only states of ferromagnetic
type. 

\begin{figure}
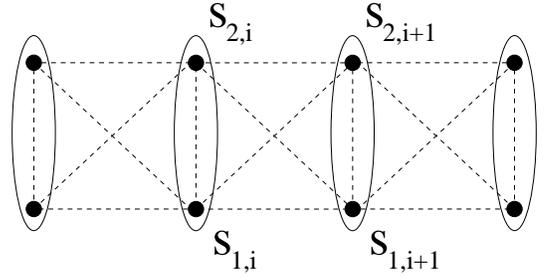

\caption{\label{fig:ladder}
Generalized spin-$1\over2$ ladder with all possible
isotropic exchange interactions between nearest-neighbor rungs.
Ovals show spin pairs whose states are involved in the elementary
matrix $g_{i}$ defined by (\protect\ref{matrix}).  
Couplings are described in the text, see (\protect\ref{ham}).}
\end{figure}

We generalize the models of
Eqs.\ (\ref{shas},\ref{aklt}) by considering the following spin
ladder Hamiltonian (see Fig.\ \ref{fig:ladder}): \\
$\widehat H = \sum_i \widehat h_{i,i+1}$, where 
\begin{eqnarray}
&& \widehat h_{i,i+1}= C_0 +\frac{1}{2}J_0 {\mathbf S}_{1,i} {\mathbf S}_{2,i}
  + \frac{1}{2}J_0' {\mathbf S}_{1,i+1} {\mathbf S}_{2,i+1} \nonumber \\
&& \; +\! J_1 {\mathbf S}_{1,i}{\mathbf S}_{1,i+1} 
        \! + \! J_1' {\mathbf S}_{2,i}{\mathbf S}_{2,i+1} 
\!+\!  J_2  {\mathbf S}_{1,i} {\mathbf S}_{2,i+1} 
    \! + \! J_2'  {\mathbf S}_{2,i} {\mathbf S}_{1, i+1} \nonumber \\
&&\; +  K ({\mathbf S}_{1,i}{\mathbf S}_{1,i+1}) 
             ({\mathbf S}_{2,i}{\mathbf S}_{2,i+1})
         \! + K' ({\mathbf S}_{1,i} {\mathbf S}_{2,i+1})  
               ({\mathbf S}_{2,i} {\mathbf S}_{1,i+1}) \nonumber \\
&& \; + K'' ({\mathbf S}_{1,i}{\mathbf S}_{2,i})
          ({\mathbf S}_{1,i+1}{\mathbf S}_{2,i+1})
\label{ham}
\end{eqnarray}
This is the general isotropic $S=\frac{1}{2}$ Hamiltonian with exchange
interactions restricted to neighboring rungs of the ladder. For periodic
boundary conditions $\frac{1}{2}(J_0  + J_0')$ is the coupling on the
rungs, in addition there are 4 pair exchange couplings on the legs and on
the diagonals and 3 biquadratic terms. Together with the constant
$C_0$ we have a total of ten parameters. One combination of these is
irrelevant since it sets the energy scale.
It is essential to include the diagonal interactions as first introduced by
White \cite{Whi96}) in order to obtain the various limits of interest.
These limits include the isotropic spin ladder and also the $S=1$ chain
when strongly ferromagnetic exchange on the rungs ($J_0 \to -\infty$) is
considered to form $S=1$ units with an effective pair interaction
$J_{eff}$ and effective biquadratic interaction $K_{eff}$
\begin{eqnarray}
J_{eff} &=&  \frac{1}{4} ( J_1 + J_1' + J_2  +  J_2' )
        + \frac{1}{8} (K + K'), \nonumber\\
K_{eff} &=& \frac{1}{4} (K + K') 
\label{Jeff}
\end{eqnarray}

Our procedure in the following is based on the ideas presented in
\cite{KluSZ91} and generalizes the work for the $S=1$ chain with biquadratic
exchange \cite{AffKLT87}: We start from the
\vskip -0.1in
\mbox{\hspace{0in}\psfig{figure=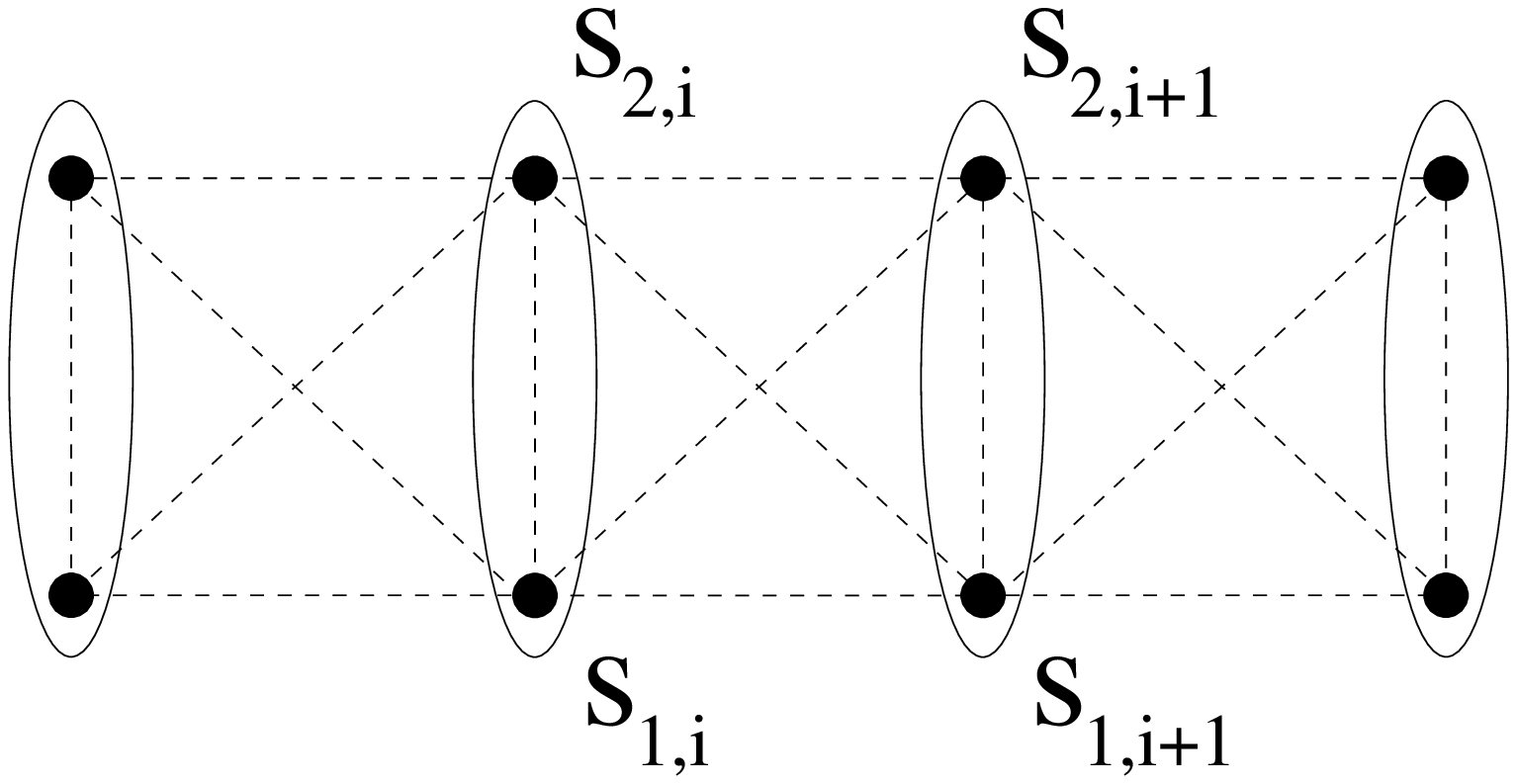,height=1.4in,angle=-90.}}
 \vskip 0.1in\nopagebreak
  \noindent\parbox[t]{3.40in}{\protect\small
FIG.\ \ref{fig:ladder}.
A generalized spin-$1\over2$ ladder with all possible
isotropic exchange interactions between nearest-neighbor rungs.
Ovals show spin pairs whose states are involved in the elementary
matrix $g_{i}$ defined by (\protect\ref{matrix}). 
Couplings are described in the text, see (\protect\ref{ham}).
}
\vskip 0.1in\noindent
MP wave function (\ref{matrix}),
considering $|s\rangle$ and $|t_{\mu}\rangle$ as the states of a single rung,
and require that the parameters in the Hamiltonian (\ref{ham}) and the free
parameter $u = b/a$ in $| \psi_0 \rangle$ satisfy the following conditions:

(i) $ | \psi_0 \rangle$ is annihilated by $H$, i.e.
\begin{equation}
\widehat h_{i, i+1} \: g_i  g_{i+1} = 0\,;
\label{cond1}
\end{equation}

(ii) all other states of $\widehat H$ have the energy $E> 0$.

\noindent
Then $ | \psi_0 \rangle$ is the ground state of $\widehat H$, with
the energy density $-C_{0}$ per rung.

To proceed we write the local Hamiltonian $\widehat h$ in the alternative
formulation \cite{NiggZitt96} using projectors on states with fixed angular
momentum of the 4 spin plaquette $(i,i+1)$ of the ladder:
\begin{eqnarray}
\widehat h &=& \lambda_{2} \sum_M  | \Psi_{2,M} \rangle \langle \Psi_{2,M} |
   + \sum_{k,l=1,2} \lambda_0^{(k,l)} | \Psi^{(k)}_{0,0} \rangle
\langle\Psi^{(l)}_{0,0} |  \nonumber\\
     &+& \sum_{k,l=1,2,3} \lambda_1^{(k,l)}
         \sum_M | \Psi^{(k)}_{1,M} \rangle \langle\Psi^{(l)}_{1,M} |
\label{proj}
\end{eqnarray}
Here $|  \Psi^{(k)}_{J,M} \rangle$  are the multiplets which can be formed
from the
4 spins $S=\frac{1}{2}$  of one plaquette: one quintuplet, three triplets
and two singlets; the 10 parameters $\lambda_J^{(k,l)} =
\lambda_J^{(l,k)}$ are linearly related to the 10 parameters in the
Hamiltonian of Eq.\ (\ref{ham}). The multiplets
$|  \Psi^{(k)}_{J,M} \rangle$ are given in terms of the singlets $|
s \rangle$ and triplets $| t \rangle$ on rungs $i$ and $i+1$; the
plaquette singlets are a linear combination of
$| ss \rangle$ and $ | (t t)_{J=0} \rangle$, the plaquette
triplets are linear combinations of $| ts \rangle$, $| st \rangle$
and $| (tt)_{J=1} \rangle$ and the plaquette quintuplet is
$| (tt)_{J=2} \rangle$. 
It is easily verified that only one singlet
\begin{equation}
(3 + u^4)^{-1/2} \left( u^2 | ss \rangle
-   \sqrt{3} | tt \rangle_{J=0} \right) \equiv
| \Psi_0^{(1)} \rangle
\label{plaq-singlet}
\end{equation}
and one triplet
\begin{equation}
(1 + u^2)^{-1/2}
\left \{ {u\over\sqrt{2}} ( | ts \rangle + | st \rangle ) - 
| tt \rangle_{J=1} \right \} \equiv | \Psi_1^{(1)} \rangle
\label{plaq-triplet}
\end{equation}
occur in $g_i g_{i+1}$. A convenient choice for the remaining
multiplets is
\begin{eqnarray}
&& | \Psi_0^{(2)} \rangle =  (3  +u^4)^{-1/2} 
\left(  \sqrt{3} | ss \rangle
+ u^2 | tt \rangle_{J=0} \right) \nonumber \\
&& | \Psi_1^{(2)} \rangle = (1/\sqrt{2}) (| ts \rangle -
| st \rangle ) \nonumber \\
&& | \Psi_1^{(3)} \rangle =
(1+u^2)^{-1/2} 
  \left\{ (1/\sqrt{2}) (| ts \rangle + | st \rangle ) +
  u | tt \rangle_{J=1} \right\} \nonumber \\
&& | \Psi_2 \rangle = | tt \rangle_{J=2}
\end{eqnarray}
Conditions (i) result in 5 equations, corresponding to
\begin{equation}
\lambda_0^{(1,1)} \:= \: \lambda_0^{(1,2)} \:= \: \lambda_1^{(1,1)} \:=
\lambda_1^{(1,2)} \:= \lambda_1^{(1,3)} \:= \:0.
\end{equation}
Thus, in order to lead to MP ground states, the Hamiltonian has to project
on the space $| \Psi_0^{(2)} \rangle, \: | \Psi_1^{(2)} \rangle, \:
| \Psi_1^{(3)} \rangle, \: | \Psi_2 \rangle$
and condition (ii) results in the inequalities
\begin{equation}
\lambda_0^{(2)} > 0, \quad \tilde{\lambda}_1^{(\alpha)} > 0, \quad
\lambda_2 > 0
\label{cond2}
\end{equation}
where $\widetilde{\lambda}_1^{(\alpha)} \: (\alpha=1,2)$ are the two
eigenvalues of the matrix $\lambda_1^{(i,j)}, \: (i,j = 2,3)$.  The
inequalities (\ref{cond2}) guarantee that the ground state is nondegenerate
(apart from the fact that Eq.\ (\ref{matrix}) describes four ground states
with different behaviour at the boundaries of open ladders), since it can be
shown by induction with respect to the ladder length that the plaquette states
of Eqs.\ (\ref{plaq-singlet},\ref{plaq-triplet}) do not allow a zero energy
state which is different from the MP ground state.

For an explicit discussion of the results we use a somewhat simplified
Hamiltonian with less freedom than in the general case: First, we set $J_0 =
J_0'$, and  from one of the
conditions (i) we find that this necessarily means $J_1 = J_1'$
(which respects the ladder symmetry) and  $\lambda_1^{(1,2)} = 0$; thus the
eigenvalues  $\widetilde{\lambda}_1^{(\alpha)}$ are identical to the
$\lambda_1^{(2,2)}$ and $\lambda_1^{(3,3)}$. 
Further we set $K - K'= K'' = 0$ (these
combinations are found irrelevant in a more detailed treatment which will
be published  separately).
Then we are left with the 6 coefficients $C_0$, $J_0$, $J_1$, $J_2$, $J_2'$,
$K$ and the conditions (i) take the following form
\begin{eqnarray}
&& 2 J_1 - J_2 - J_2' \:+ \: u^2 \left( 4C_0/3 - J_0 + K/2
 \right) = 0 \nonumber \\
&& 2(2J_1 +  J_2 +  J_2') - 4C_0 - J_0 - 7K/2  \\
\label{cond}
&&\qquad\qquad\qquad\qquad\qquad\qquad  + u^2 (J_2 + J_2' - 2 J_1) = 0
\nonumber \\ 
&&  2 (J_2'-J_2) + u \left( 4 C_0 - J_0 + 2 J_1 - J_2 - J_2'
-  K/2 \right) = 0 \nonumber \\
&& 2 J_1 + J_2 + J_2' + 3 K/2 - 4 C_0 - J_0
  +  2 u (J_2 - J_2')  = 0 \nonumber
\end{eqnarray}
The general solution of these 4 linear equations contains 2 arbitrary
constants, when the 
parameter $u = b/a$ of the MP wavefunction is fixed. We
absorb one of the two constants in the energy scale, denote the remaining
one by $x$ and define
\[
u^2 = F
\]
to obtain the following family of hamiltonians with exact MP
eigenstate:
\begin{eqnarray}
&& C_0 = 3 \left( 9 +2 x - 3 (1 + x)F + 3 x F^2  \right)/32\,,
\nonumber \\
&& J_0 = 3 \left( 2 - x - (1 + x)F + x F^2 \right)/4\,,
      \nonumber \\
&& 2 J_1 + J_2 + J_2' = \left(15 - 9  (1 + x)F + 7 x F^2
    \right)/4 \,, \nonumber \\
&& 2 J_1 - J_2 - J_2' = - x F\,,  \nonumber \\
&& 2 J_2 - 2 J_2' = F^{1/2} \left( 3(1+ x)/2
     -  x F \right)\,,  \nonumber  \\
&& K = 3 \left( F - 1 \right) \left( x F -1 \right)/4\,.
\end{eqnarray}
With these definitions the conditions of Eq.\ (\ref{cond2})
that the MP state is a ground state are obtained
as follows
\begin{eqnarray}
&& \lambda_0 =  x (3 + F^2(x))/4 \:> \:0 \nonumber \\
&& \widetilde\lambda^{(1)}_1  =  \left( 3 (1+x)  + 2x F(x)
     \right)/8  \: > \: 0  \nonumber \\
&& \widetilde\lambda^{(2)}_1  =  (1 + F(x)) \left( 3 (1+x)
     - 2x F(x)/8 \right)  \: > \: 0   \nonumber \\
&& \lambda_2  =  \left( 18 + 8 x F^2(x)  
  - 9 (1+x) F(x) \right)/8 \:  > \: 0
\label{lambdas}
\end{eqnarray}
The eigenvalues $\lambda_0$ and $\widetilde\lambda_1^{(1)}$ are
seen to be positive for $x>0$, $F>0$.  The eigenvalues
$\widetilde\lambda_1^{(2)}$ and
$\lambda_2$ require a more
detailed discussion; 
it is easily seen that a sufficient (though not necessary) condition for
$\lambda_2 > 0$ is $J_{eff} > 0$, i.e. an effective antiferromagnetic
interaction between the two rungs of one plaquette, and the condition on
$\widetilde \lambda^{(2)}_1$ is equivalent to the requirement
$u(J_{2}-J_{2}')>0$. 
It is convenient to discuss the possible hamiltonians considering $F$ as
function of $x$. We use $F=u^{2}$ instead of $u$ keeping in mind that
changing the sign $u\to-u$ amounts just to interchanging the ladder legs and
thus does not bring in any new physics; therefore from now on we assume that
$F^{1/2}>0$. 

The following illustrative members of this family of hamiltonians
are now easily obtained:

(i) Shastry-Sutherland model:
\begin{equation}
F= 1, \qquad x = 3\delta/ (2 + \delta),\quad 0<\delta<1 \,.
\label{SS}
\end{equation}
Since the MP wave function with $F=1$, i.e. $a=b$ corresponds
to singlets on one type of diagonals \cite{BreMN96}, this also covers
(after a translation of one of the ladder legs) the case $a=0$.
The eigenvalue $\lambda_0$ vanishes for $x=0$, which
is to be expected since in the Majumdar-Ghosh limit a second degenerate
eigenstate exists (singlets on the alternative bonds).

The solution (16) also applies for partially ferromagnetic
interactions, $x>1$, resp. $(1-\delta)/(1+\delta) < 0 $. Although the
effective interaction between units on the diagonals is ferromagnetic,
the Shastry-Sutherland dimer state is the ground state up to $x=9$
(corresponding to $\delta = -3$), since all $\lambda$s are
positive. At the singular point $\delta=-3$ the eigenvalue $\lambda_2$
vanishes and the energy of the dimer ground state coincides with the
energy of the fully polarized ferromagnetic configuration. Thus this is
the point of the first-order quantum phase transition from dimer to
ferromagnetic phase.

(ii) Generalized AKLT model, defined by $F=0$, i.e. $b=0$ in the MP wave
function (so that only triplets on the rungs occur):
\begin{equation}
F=0, \qquad x \: \:{\rm finite}
\end{equation}
When choosing a convenient scale factor to render $J_{eff} = 1$ this
corresponds to the following Hamiltonian:
\begin{eqnarray}
&& J_0 = 4/3 - 2x/3,\;\;
J_1 = J_2 = J_2' = 5/6, \nonumber\\
&& K = K' = 2/3,\;\; K'' = 0,\;\; C_0 = 3/4 + x/3 .
\end{eqnarray}
This is essentially the Hamiltonian of Ref.\ \onlinecite{AffKLT87}, but
without requiring explicitly the coupling of two spins $S=\frac{1}{2}$ into
a triplet, a result that could be easily obtained directly. The AKLT model in a
strict sense is obtained for $x \to +\infty$ and $x \:F(x) \: \to \: 0 $.

(iii) For $x F(x)  \:= \: 1$ another class of hamiltonians {\em without
biquadratic terms} is
obtained; the condition $\lambda_{2}>0$ leads to the restriction $x\geq
{1\over9}$.  A nontrivial case is 
$x = \frac{1}{9}$, which leads to the presence of only one diagonal
exchange interaction, equivalent  to a chain with NN and NNN exchange,
\begin{eqnarray}
&& C_0 = 7/12, \; J_0 = J_0' = 2/3, \;
J_1 = J_1' = J_2' =-1,\nonumber\\
&& J_2 = K = K' = K'' = 0.
\end{eqnarray}
This is a $S=\frac{1}{2}$ chain with alternating ferro- and antiferromagnetic
exchange and ferromagnetic NNN interactions, a frustrated Heisenberg chain. As
in (i), the eigenvalue $\lambda_2$ vanishes and the energy of the fully
polarized ferromagnetic state coincides with that of the MP ground state
(\ref{matrix}). Thus we have obtained two exact points on the line where the
chain with NN and NNN interactions \cite{BreMN96} undergoes a first order
quantum phase transition to the ferromagnetic state.

(iv) A family of models with exact MP ground states connecting the
Majumdar-Ghosh and the AKLT limits is obtained from e.g.
\begin{equation}
F(x) \:= \:  1/(1+x)^2.
\label{inter}
\end{equation}
Since the Majumdar-Ghosh and dimer chains have identical ground states,
Eq.\ (\ref{inter}) explicitly demonstrates the possibility to transform the
Hamiltonian of the dimer $S=\frac{1}{2}$ chain
continuously to the Hamiltonian of the AKLT chain without any
singularity (or quantum phase transition) in the ground state. 

This provides an exact proof that from the point of view of theoretical
concepts no essential difference exists between the Haldane gap and the
dimer gap and that the gapped $S=\frac{1}{2}$ models listed in the
introduction belong to the same phase as the Haldane
$S=1$ chain.

The authors gratefully acknowledge financial support from Deutsche
Forschungsgemeinschaft (AK) and Volkswagenstiftung (HJM) and the hospitality
of the Hannover Institute for Theoretical Physics (AK) and the Department of
Earth and Space Science at Osaka University (HJM)). AK was also supported by
the grant 2.4/27 from the Ukrainian Ministry of Science.

\begin{references}

\bibitem[\ast]{perm1} Permanent address: Institute of Magnetism,
36(b) Vernadskii av., 252142 Kiev, Ukraine.

\bibitem[\dagger]{perm2} Permanent address: Institut f\"ur Theoretische Physik,
Universit\"at Hannover, 30167 Hannover, Germany.

\bibitem{Bet31} H. A. Bethe, Z. Physik {\bf 71}, 205 (1931).

\bibitem{Hal83} F. D. M. Haldane, Phys. Rev. Lett. {\bf 50}, 1153 (1983).

\bibitem{DagR96} E. Dagotto and T. M. Rice, Science {\bf 271}, 618 (1996).

\bibitem{NagTCPS91} S. E. Nagler, D. A. Tennant, R. A. Cowley,
T. G. Perring and S. K. Satija, Phys. Rev. B {\bf 44}, 12361 (1991).

\bibitem{ZheSSHU96} A. Zheludov, G. Shirane, Y. Sasago, M. Hase and G.
Uchinokura, Phys. Rev. B {\bf 53}, 11642 (1996).

\bibitem{EccBBJ94} R. S. Eccleston, T. Barnes, J. Brody and
J. W. Johnson, Phys. Rev. Lett. {\bf 73}, 2626 (1994).

\bibitem{AzuHTIK94} M. Azuma, T. Hiroi, M. Takano, K. Ishida and Y.
Kitaoka, Phys. Rev. Lett. {\bf 73}, 3463 (1994).

\bibitem{RenGRV90} J. P. Renard, V. Gadet, L. P. Regnault and
M. Verdaguer, J. Magn. Magn. Mat. {\bf 90-91}, 213 (1990).

\bibitem{KohT92} M. Kohmoto and H. Tasaki, Phys. Rev. B {\bf 46},
3486 (1992).

\bibitem{Hid91} K. Hida, J. Phys. Soc. Jpn. {\bf 60}, 1347 (1991);
{\bf 64}, 4896 (1995).

\bibitem{KatI95} N. Katoh and M. Imada, J. Phys. Soc. Jpn. {\bf 63},
1437 (1995).

\bibitem{TotS95a} K. Totsuka and M. Suzuki, J. Phys.: Condens. Matter {\bf
7}, 6079 (1995).

\bibitem{BreMN96} S. Brehmer, H.-J. Mikeska and U. Neugebauer,
J. Phys: Condens. Matter {\bf 8}, 7161 (1996).

\bibitem{Whi96} S. R. White, Phys. Rev. B {\bf 53}, 52 (1996).

\bibitem{ShaS81} B. S. Shastry and B. Sutherland, Phys. Rev. Lett.
{\bf 47}, 964 (1981).

\bibitem{MajG69} C. K. Majumdar and D. K. Ghosh, J. Math. Phys. {\bf 10},
1399 (1969).

\bibitem{AffKLT87} I. Affleck, T. Kennedy, E. Lieb and H. Tasaki, Phys.
Rev. Lett. {\bf 59}, 799 (1987); Commun. Math. Phys.
{\bf 115}, 583 (1988). 2578 (1987).

\bibitem{FanNW89} M. Fannes, B. Nachtergaele and R. F. Werner,
Europhys. Lett. {\bf 10}, 633 (1989); 
Commun. Math. Phys. {\bf 144}, 443 (1992).

\bibitem{KenT92} T. Kennedy and H. Tasaki, Phys. Rev B {\bf 45},
304 (1992).

\bibitem{KluSZ91} A. Kl\"umper, A. Schadschneider and J. Zittartz, J. Phys.
A {\bf  24}, L955 (1991);
Europhys. Lett. {\bf 24}, 293 (1993).

\bibitem{NiggZitt96}
H. Niggemann and J. Zittartz, Z. Physik B {\bf 101}, 289 (1996).

\bibitem{GSu96} Gang Su, Phys. Lett. A {\bf 213}, 93 (1996).)

\bibitem{TotS95} K. Totsuka and M. Suzuki, J. Phys.: Condens. Matter {\bf
7}, 1639 (1995).

\bibitem{KolMY97} A. K. Kolezhuk, H.-J. Mikeska and Shoji Yamamoto, 
to appear in Phys. Rev. B {\bf 55} (1997).

\end{references}
\end{document}